\documentclass[twocolumn]{jpsj2} 
\title{Magnetic Fluctuations in the Metallic State of Na$_{0.7}$CoO$_2$ Revealed by $^{23}$Na NMR}

\author{
Y. \textsc{Ihara,}$^{1}$\thanks{E-mail address: ihara@scphys.kyoto-u.ac.jp}
K. \textsc{Ishida,}$^{1}$\thanks{E-mail address: kishida@scphys.kyoto-u.ac.jp} 
C. \textsc{Michioka,}$^{2}$
M. \textsc{Kato,}$^{2}$
K. \textsc{Yoshimura,}$^{2}$
H. \textsc{Sakurai,}$^{3}$
and
E. \textsc{Takayama-Muromachi}$^{3}$}

\inst{$^{1}$Department of Physics, Graduate School of Science, Kyoto University, Kyoto 606-8502, Japan. \\
$^{2}$Department of Chemistry, Graduate School of Science, Kyoto University, Kyoto 606-8502, Japan. \\
$^{3}$Superconducting Materials Center, National Institute for Materials Science, 1-1 Namiki, Tsukuba, Ibaraki 305-0044, Japan.}

\abst{Nuclear-magnetic-resonance (NMR) studies were performed in Na$_{0.7}$CoO$_2$ to investigate its magnetic properties from the microscopic point of view. Na$_{0.7}$CoO$_2$ is a starting compound of the recently discovered superconductor Na$_{0.35}$CoO$_2$$\cdot 1.3$H$_2$O, in which Co atoms form a triangular structure. Although $^{23}$Na-NMR results show no magnetic anomaly down to 1.5 K, Knight shift ($^{23}K$) and $^{23}(1/T_1T)$ of $^{23}$Na continue to increase, suggestive of the vicinity of the magnetic phase. From the relation between $^{23}K$ and $^{23}(1/T_1T)$, it is shown that ferromagnetic fluctuations are dominant below 100 K. In addition to the ferromagnetic correlations, antiferromagnetic fluctuations are also developed in a low temperature region below 4 K. Magnetic fluctuations suggested by our experiment are discussed.
}

\kword{superconductivity, hydrous sodium cobalt oxide, NQR, spin fluctuations}

\begin{document}
\maketitle

Since the superconductivity in Na$_x$CoO$_2$$\cdot y$H$_2$O ($x \sim 0.35$, $y \sim 1.3$) with the transition temperature $T_c \sim 5$ K was discovered by Takada {\it et al}\cite{Takada1}, understanding of magnetic properties in the CoO$_2$ plane is one of the most important topics in strongly correlated electron systems.
This is because the magnetic properties in the plane might be related to the mechanism of the unconventional superconductivity in the compound. In addition, the Co atoms form a triangular lattice in the plane, which is anticipated to possess the geometrical frustrations. Novel types of magnetic transitions and ground states are expected due to a suppression of a long-range magnetic ordering by the geometrical frustrations associated with the triangular structure. In particular, only a few compounds have been known to date which possess a metallic property down to the lowest temperature with the geometrical frustrations, e.g. (YSc)Mn$_2$\cite{YMn2} and LiV$_2$O$_4$\cite{LiV2O4}. The Na$_x$CoO$_2$ ($0.5 < x < 0.78$) compound is considered to be one of the promising metallic compounds with the geometrical frustrations at low temperature\cite{Sakurai}.    

In this paper, we report the magnetic properties in Na$_{0.7}$CoO$_{2}$ revealed by our $^{23}$Na-NMR experiments.
Powder samples of Na$_{0.7}$CoO$_2$ were prepared from Na$_2$CO$_3$ (99.99\%) and Co$_3$O$_4$(99.9\%) by solid-state reaction. Its quality was checked by X-ray powder diffraction.  The Na content, which is very sensitive to physical properties, is estimated to be 0.702 by the inductive-coupled plasma atomic emission spectroscopy (ICP-AES) method. Using this well-characterized powder sample, $^{23}$Na- and $^{59}$Co-NMR measurements were performed.

\begin{figure}[tb]
\begin{center}
\includegraphics[width=7cm]{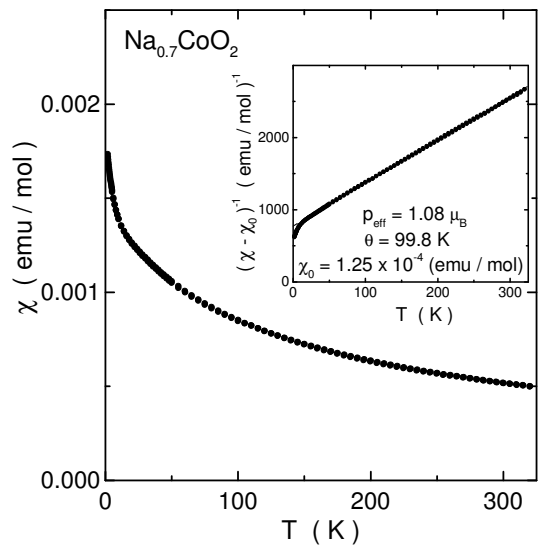}
\end{center}
\caption{Temperature dependence of bulk susceptibility of $\chi$ in 10 kOe. The inset shows a plot of the inverse of $\chi$-$\chi_0$ against $T$.  The parameters fitted by $\chi(T)-\chi_0 = N_{A}P_{\rm eff}^2/(T+\theta)$ are shown in the inset.Deviation from the CW behavior is seen below 25 K.}
\label{f1}
\end{figure}

The main panel of Figure 1 shows temperature dependence of magnetic susceptibility $\chi$ in 10 kOe. $\chi$ increases with decreasing temperature, following the Curie-Weiss (CW) behavior. The inset of Fig.~1 shows the inverse magnetic susceptibility $(\chi-\chi_0)^{-1}$ as a function of temperature. Fairly good CW type behavior is observed above 50 K. Fitting parameters by the CW formula are shown in the inset. These values are in good agreement with previous reports\cite{Miyoshi,Gavilano}. As pointed out by Gavilano {\it et al.}\cite{Gavilano}, the size of the effective moment can be accounted for by the mixed-valence configurations that Co$^{4+}$ ($S$ = 1/2) and Co$^{3+}$ ($S$ = 0) are the ratio of 0.38 : 0.62. This ratio is not so inconsistent with the mixed-valence state suggested from the Na content (Co$^{4+}$:Co$^{3+}$=0.3 : 0.7).

\begin{figure}[tb]
\begin{center}
\includegraphics[width=8cm]{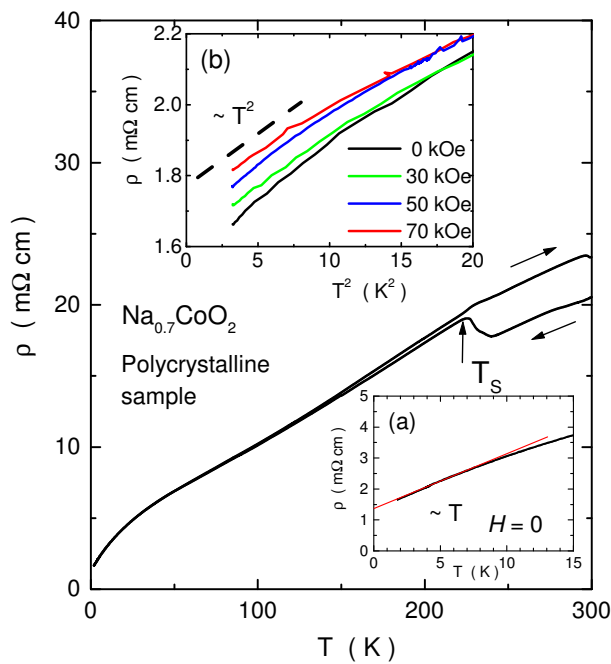}
\end{center}
\caption{Temperature dependence of resistivity $\rho$ in polycrystalline sample below 300 K measured in zero field. The insets of (a) and (b) show the $T$ dependence of $\rho$ below 10 K, and $T^2$ plot in various fields, respectively.}
\label{f1}
\end{figure}

Figure 2 shows the temperature dependence of resistivity $\rho$ measured by the polycrystalline Na$_{0.7}$CoO$_2$. Above 220 K, $\rho$ shows a hysterisis behavior, which suggests a first order transition. Quite recently, neutron powder diffraction measurements revealed that structual transition ascribed to Na rearrangement occurs in Na$_x$CoO$_2$ with $x \sim 0.7$\cite{Narearrangement}. The anomaly in $\rho$ above $T_S \sim 220$ K is due to this crystal anomaly. The resistivity shows metallic behavior below 200 K, especially, $\rho$ is proportional to $T$ below 8 K as shown in the inset (a) of Fig.~2. It seems that the application of a magnetic field leads to a gradual recovery of a $T^2$ behavior as shown in the inset (b) of Fig.~2. These resistivity results remind us of the non-Fermi-liquid behavior seen in the heavy-fermion system\cite{NFLCeCu6Au}. Resistivity measurement using a high-quality single crystal is, however, needed for the fully identification of the intrinsic behavior of $\rho$\cite{rhocomment}. We point out that the mixed-valence state suggested by $\chi$ is inconsistent with this metallic behavior because the resistivity would show the insulating behavior in such the ionic mixed-valence state. In order to investigate magnetic properties in the low-temperature metallic state, we performed NMR measurements on the same polycrystalline sample in which $\chi$ and $\rho$ were measured. 

\begin{figure}[tb]
\begin{center}
\includegraphics[width=8cm]{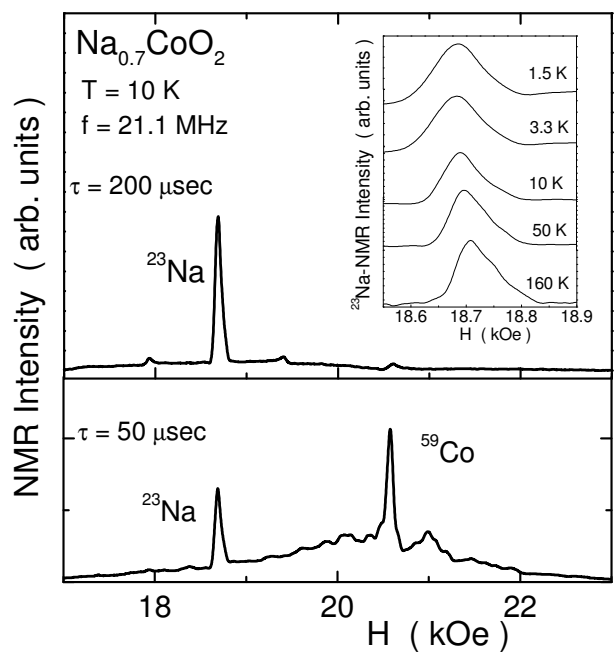}
\end{center}
\caption{Na and Co NMR spectra in Na$_{0.7}$CoO$_2$ obtained at 21.1 MHz and at 10 K. The NMR spectrum obtained at $\tau = 50 \mu$sec (bottom) and 200 $\mu$sec (upper) are shown, where $\tau$ is a time interval between two pulses for the NMR spin-echo measurement. The inset shows the central peak of Na spectrum corresponding to $1/2 \leftrightarrow -1/2$ transition measured in various temperatures.  }
\label{f1}
\end{figure}

NMR spectra were measured at a fixed frequency of 21.1 MHz using the randomly oriented powdered sample. Figure 3 shows $^{59}$Co- and $^{23}$Na-NMR spectra, in which the spin-echo intensity was recorded during sweeping magnetic field. First, we measured the $^{59}$Co-NMR quantities since its signal is so strong as to be observed even at room temperature. The $^{59}$Co Knight shift $^{59}K$ was measured at the sharp peak ($H_{res} \sim$ 20.6 kOe in Fig.~3) in the $^{59}$Co-NMR spectrum. The $^{59}$Co-Knight shift shows a weak temperature dependence up to 50 K and is nearly constant above 50 K. The temperature dependence of $^{59}K$ is not scaled at all with $\chi$. It seems that NMR quantities obtained from the sharp peak in the Co-NMR spectrum is not directly affected by the Co-$3d$ electrons, thus we performed Na NMR measurements instead. The $^{59}$Co-NMR results will be summarized in a separated paper.      

Although the $^{23}$Na- and $^{59}$Co-NMR signals are overlapped as seen in Fig.~3 when we recorded the spin-echo signal at $\tau = 50 \mu$sec, the Co NMR spectrum becomes undetectably small and we can observe only Na spectrum when $\tau$ is 200 $\mu$sec as shown in the upper figure of Fig.~3. Here $\tau$ is the time interval between two pulses by means of spin-echo method. This is because the nuclear spin-spin relaxation time ($T_2$) of the Co site is so short that NMR spin-echo intensity $I$ is negligibly small at $\tau = 200 \mu$sec since $I(\tau)$ decreases following exp($-2\tau/T_2$). We adopt $\tau = 200 \mu$sec for the Na-NMR measurement to avoid the effect from the Co-NMR signal. The Na NMR spectrum with $\tau$ = 200 $\mu$sec consists of an intense peak arising from $1/2 \leftrightarrow -1/2$ transition, two small satellite peaks arising from $\pm 1/2 \leftrightarrow \pm 3/2$ transitions and broad background signal over a few kOe. From the split of the two satellite peaks, the electric field gradient (EFG) frequency was estimated to be 1.71 MHz. This value is in good agreement with the EFG frequency at ``Na1'' site reported by Mukhamedshin {\it et al.}\cite{Mukhamedshin}, where the ``Na1'' site is the (0,0,1/4) site which is vertically aligned to two Co atoms. When we compare this Na-NMR spectrum to a typical powder-pattern NMR spectrum, we notice that the peak of the central transition in our spectrum is much more intense than that expected from the satellite peaks.\cite{MetallicShift} It is considered that the Na signal from other site overlaps in the central-transition peak. According to the paper\cite{Mukhamedshin}, a Na-NMR signal from ``Na2'' site at (2/3,1/3,1/4), which is the center of Co triangle in the neighboring CoO$_2$ planes, are identified using an aligned powder sample under strong magnetic field. It was revealed that the central peak of the Na2-NMR spectrum is very close to that of the Na1-NMR spectrum. Therefore, it is considered that the central peaks of the Na1 and Na2-NMR spectra are overlapped in our randomly oriented powder sample. The inset in Fig.~3 shows the overlapped central peak at various temperatures. Appreciable splitting of the Na1- and Na2-peaks was not observed in the measured temperature range. This is consistent with the previous experimental results that temperature dependence of $^{23}K$ at Na1 and Na2 sites are the same, since the central peaks of the Na1- and Na2-NMR spectra shift in the same way. Therefore the Knight shift estimated from the overlapped central peak in our powder sample is regarded as an isotropic component of the averaged Knight shift of the Na1 and Na2 sites.

\begin{figure}[tb]
\begin{center}
\includegraphics[width=8cm]{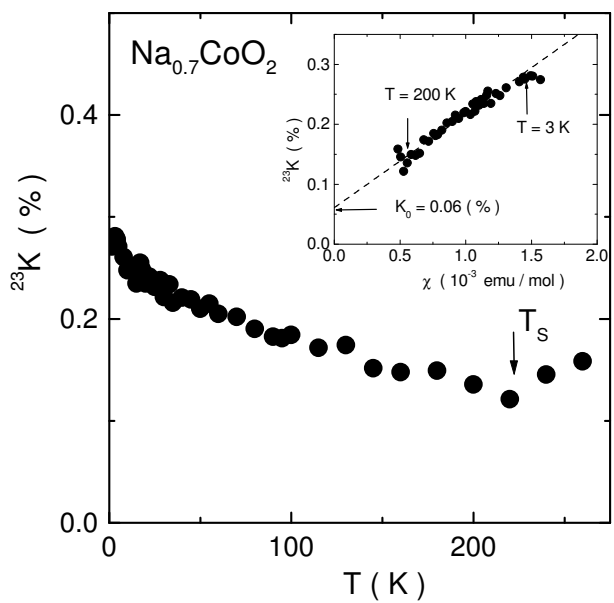}
\end{center}
\caption{$T$ dependence of Knight shift of $^{23}$Na $^{23}K$ derived from the central NMR spectrum shown in the inset of Fig.~3. The inset shows the plot of the $^{23}K$ against $\chi$ with temperature as an implicit parameter.}
\label{f1}
\end{figure}

Figure 4 shows the temperature dependence of the $^{23}K$.
The $^{23}$Na-Knight shift increases continuously with decreasing temperature, and the temperature dependence is quite similar to that of $\chi$.
The inset of Fig. 4 shows the plot of the $^{23}K$ as a function of $\chi$. As seen in the figure, a good relation holds between two quantities in the temperature range between 3 and 200 K. The deviation above 200 K is due to the structural anomaly seen in $\rho$. The good linear relation between $^{23}K$ and $\chi$ implies that the microscopic susceptibility at the Na sites are governed by the bulk susceptibility originating from the Co-$3d$ electrons. It is noteworthy that the upturn of $\chi$ below 20K is not due to some impurity phase, but an intrinsic behavior.  From the linear relation between $^{23}K$ and $\chi$ between 3 and 200 K, the hyperfine coupling constant and the intercept of the $y$ axis are evaluated to be $8.69 \pm 0.44$ kOe/$\mu_B$ and $K_0 = 0.061 \pm 0.007$ \%. Taking into account that this shift corresponds to the isotropic term of the Knight shift, the hyperfine coupling constant arises from the covalency between Na-$2s$ and Co-$3d$ orbitals by way of the O atom. The temperature independent term of the $^{23}K$ is considered to arise mainly from the EFG effect at the Na site since $^{23}K$ does not have an orbital term in most cases.     

\begin{figure}[tb]
\begin{center}
\includegraphics[width=8cm]{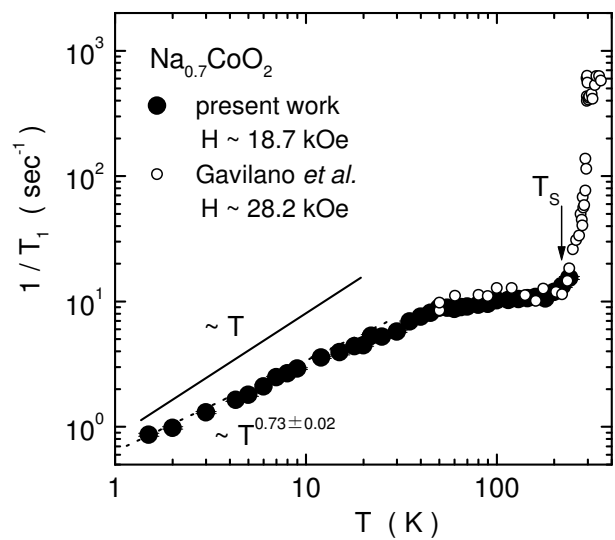}
\end{center}
\caption{$T$ dependence of $1/T_1$ measured at the Na-NMR central peak. $1/T_1 $ above 50 K measured by Gavilano {\it et al.} is shown for comparison. $1/T_1$ below 10 K approximately follows $T^{0.73 \pm 0.02}$ relation.  }
\label{f1}
\end{figure}

Figure 5 shows the temperature dependence of $1/T_1$ of Na $^{23}(1/T_1)$. $^{23}T_1$ was determined by fitting the nuclear magnetization $m(t)$ after the saturation to the theoretical curve for the central transition of $I=3/2$. $^{23}T_1$ was evaluated to be a unique value in the whole temperature range. As shown in the figure, our $1/T_1$ result above 50 K is in good agreement with the previous report by Gavilano {\it et al.}\cite{Gavilano}, although we did not measure $T_1$ above 250 K. $1/T_1$ shows a clear anomaly around room temperature, which is due to a structural anomaly seen in $\rho$ and $^{23}K$. With decreasing temperature, $1/T_1$ is nearly constant down to 50 K, and starts to decrease below 50 K. It is noteworthy that the resistivity $\rho$ shows a steep decrease around 40 K. It is considered that some change in the electronic states occurs around this temperature. In the low temperature range, $1/T_1$ seems to follow a $T^n$ relation below 10 K. From the fitting of the data, $n$ is estimated as $n = 0.73 \pm 0.02$. The $T_1T$ = constant relation, which is the characteristic feature of the Fermi-liquid state, is not observed down to 1.5 K, suggestive of the existence of the strong magnetic fluctuations at low temperatures.

\begin{figure}[tb]
\begin{center}
\includegraphics[width=8cm]{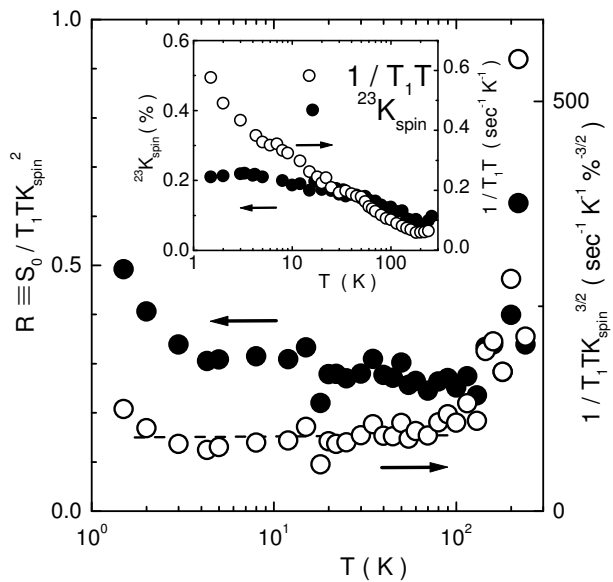}
\end{center}
\caption{$T$ dependence of $R \equiv S_0/T_1TK_{\rm spin}^2$ (left axis)  and $1/T_1TK^{3/2}$ (right axis) (see in the text). The inset shows $T$ dependence of $^{23}K_{\rm spin}$ (left axis) and $1/T_1T$ (right axis).}
\label{f1}
\end{figure}

Now, we discuss the spin-fluctuation character at low temperature region in Na$_{0.7}$CoO$_2$ from the results of the spin part of $^{23}K$ and $^{23}(1/T_1T)$. When NMR shift and $1/T_1T$ arise from $s$ electrons as in the present case, the Korringa relation is known to be satisfied when the electronic interaction is negligibly small. The noninteracting Korringa relation is expressed by ($T_1TK_{\rm spin}^2)_0 = (\gamma_e/\gamma_n)^2(\hbar/4\pi k_B) \equiv S_0$, where $\gamma_e$ and $\gamma_n$ are the electronic and nuclear gyromagnetic ratios. In the presence of interactions, the ratio $R \equiv S_0 / T_1TK_{\rm spin}^2$ provides an important information about magnetic correlations. Since $1/T_1T$ is related to the low-energy part of the $q$-dependent dynamical susceptibility in compounds, it can be enhanced by either ferromagnetic (FM) and antiferromagnetic (AFM) spin correlations, while only FM correlations strongly enhance the spin shift. Therefore, the value of $R$ is much small (larger) than unity when FM (AFM) correlations become significant. Using the experimental values of $^{23}K_{\rm spin}$ and $^{23}(1/T_1T)$, $R$ is estimated as shown in the main figure of Fig.~6. $R$ enhances above 200 K due to the additional dynamics associated with the Na rearrangement.
Below 150 K, $R$ remains at a small value of $R \sim 0.3$ down to 4 K, which points to a predominance of FM correlations. The FM correlations revealed by NMR are consistent with the larger Wilson ratio $R_W \sim 2.8$\cite{Sakurai}.

In addition, when we plot experimental data of $1/T_1TK^{3/2}$ against $T$ as shown in Fig.~6, we found that $1/T_1TK_s^{3/2}$ is nearly constant up to 100 K within our experimental accuracy. This constant relation implies a linear relation between $1/T_1T$ and $\chi_{\rm spin}^{3/2}$. According to the self-consistent renormalization (SCR) theory for weak or nearly FM metals\cite{SCR}, it was predicted that $1/T_1T$ is proportional to $\chi_{\rm spin}$ and $\chi_{\rm spin}^{3/2}$ under the predominance of three-dimensional (3-D) and 2-D FM correlations, respectively. The relation of $1/T_1T \propto \chi_{\rm spin}^{3/2}$ and the small value of $R$ suggest that a 2-D FM correlations are dominant in Na$_{0.7}$CoO$_2$. Quite recently, it was reported from a neutron scattering (NS) experiment that a prominent inelastic signal at low energies is localized around $Q_{2D}$ = (0,0) in the 2-D reciprocal space\cite{NS}. The observation of this peak is a direct evidence for the existence of the FM spin fluctuations in the CoO$_2$ layers. Our results are in good agreement with this NS result.

We also point out that single component of $^{23}(1/T_1)$ down to 1.5 K suggests a homogeneous magnetic state. If the mixed valence state suggested from the $\chi$ result\cite{Gavilano,Mukhamedshin}, $1/T_1$ would have two components at low temperatures. Obviously, it is not a case. Our $^{23}(1/T_1)$ result strongly suggests that the magnetic properties in the CoO$_2$ plane should be understood by a homogeneous itinerant magnetic state. Therefore, the deviation from the CW behavior below 20 K is considered to be due to the development of the FM correlations.

In addition, it is noteworthy that $1/T_1T$ continues to increase down to 1.5 K although $^{23}K_{\rm spin}$ levels off below 4 K. The $R$ increases below 4 K but is still less than unity at 1.5 K. This disparate behavior between $1/T_1T$ and $^{23}K_{\rm spin}$ at low temperatures shows that the AFM correlations also develop below 4 K in the predominance of the FM correlations. Our Na-NMR study suggests that the FM and AFM correlations compete with each other at low temperatures. The competing magnetic fluctuations would be related to the Fermi surface (FS) of Na$_{0.7}$CoO$_2$. According to the band calculation,\cite{Singh} two different kinds of FS's are shown: small six hole pockets near the K point which have an $e'_g$-orbital character, and large cylindrical hole FS which has a dominant $a_{1g}$-orbtal character. It is considered that the former FS's favor the FM fluctuations and the latter FS would induce the AF ones. We suggest that the these FS's with the different character are essential to understand the magnetic properties in Na$_{0.7}$CoO$_2$.            

In conclusion, the 2-D FM correlations are shown from the small Korringa ratio of $R \sim 0.3$ and the relation of $1/T_1T \propto \chi_{\rm spin}^{3/2}$. Our NMR results are consistent with the large Wilson ration $R_W \sim 2.8$ and the recent NS result. In addition to this FM correlations, the AFM correlations are also shown from the different behaviors between $1/T_1T$ and $K_{\rm spin}$ below 4 K. We suggest that the competing magnetic fluctuations at low temperaturs are due to the two hole FS's in Na$_{0.7}$CoO$_2$, which have different characters each others.

We thank Y.~Maeno, H.~Yaguchi and S.~Nakatsuji for experimental supports and valuable discussions. We also thank H.~Ikeda, S.~Fujimoto, K.~Yamada, Y.~Yanase, M.~Ogata for valuable discussions.
This work was supported by the 21 COE program on ``Center for Diversity and Universality in Physics'' from MEXT of Japan, and by Grants-in-Aid for Scientific Research from the Japan Society for the Promotion of Science (JSPS) and MEXT.


\begin{thebibliography}{99} 


\bibitem{Takada1}
K.~Takada {\it et al.}, Nature (London) {\bf 422} (2003) 53.


\bibitem{YMn2}
M.~Siga {\it et al.}, J. Phys. Soc. Jpn. {\bf 62} (1993) 1329.


\bibitem{LiV2O4}
S.~Kondo {\it et al.}, Phys. Rev. Lett. {bf 78} (1997) 3729.


\bibitem{Sakurai}
H.~Sakurai {\it et al.}, unpublished.


\bibitem{Miyoshi}
K.~Miyoshi {\it et at}, cond-mat/0403028.


\bibitem{Gavilano}
J.~L.~Gavilano {\it et al.}, Phys. Rev. B {\bf 69} (2004) 100404.


\bibitem{Narearrangement}
Q. Huang {\it et al.}, cond-mat/0405677.


\bibitem{NFLCeCu6Au}
H. v. Lohneysen, J. of Phys. Condens. Matter 8, 9689 (1996).

\bibitem{rhocomment}
Recent Li {\it et al.} (cond-mat/0401099) reported that $\rho$ measured in a single crystal of Na$_{0.7}$CoO$_2$ shows the $T^2$ behavior below 1 K in a zero field and that an application of the magnetic field makes the temperature region of the $T^2$ behavior wider. The single-crystal result is qualitativily consistent with our polycrystalline results.  


\bibitem{Mukhamedshin}
I.~R.~Mukhamedshin {\it et al.}, cond-mat/0402074.


\bibitem{MetallicShift}
G.~C.~Carter {\it et al.} ``Metallic Shift in NMR '' (Pergamon Press, Oxford 1977).

\bibitem{SCR}
T.~Moriya, and K.~Ueda, Solid State Commun. {\bf 15} (1974) 169.

\bibitem{NS}
A.~T.~Boothroyd, {\it et al.}, cond-mat/ 0312589.

\bibitem{Singh}
D.~J.~Singh, Phys. Rev. B {\bf 61} (2000) 13397.


\end{thebibliography}
\end{document}